\documentclass[rnote]{aa}  
\usepackage{graphicx,natbib,ulem}
\usepackage{txfonts}
\usepackage{color}
\begin{document}
  \titlerunning{The circumbinary disk of SS 433}
\authorrunning{ Bowler}
   \title{Yet more on the circumbinary disk of SS 433 }

   \subtitle{}

   \author{M. G.\ Bowler \inst{}}

   \offprints{M. G. Bowler \\   \email{m.bowler1@physics.ox.ac.uk}}
   \institute{University of Oxford, Department of Physics, Keble Road,
              Oxford, OX1 3RH, UK}
   \date{Received; accepted}

% \abstract{}{}{}{}{} 
% 5 {} token are mandatory
 
  \abstract
  % context heading (optional)
  % {} leave it empty if necessary  
   {Certain  lines in spectra of the Galactic microquasar SS\,433, in particular  the H$\alpha$ line and He I, have been interpreted as emission from a circumbinary disk. In this interpretation the orbital speed of the glowing material is in excess of 200 km s$^{-1}$ and the mass of the binary system in excess of 40$M_\odot$. The data show that He I emission fades much faster than H$\alpha$. This requirement has been incorporated into successful heuristic models yet has remained unexplained hitherto.} 
  % aims heading (mandatory)
  {To present a model in which the disk is excited by radiation from the close environment of the compact object and in which the different characteristics of the H$\alpha$ and He I lines are naturally explained by the different ionization potentials of the atoms.}
  % methods heading (mandatory)
   { A model in which the emission of any given patch of putative circumbinary disk material is determined  both by its illumination and the depth of the ionization zone was constructed. The illumination is proportional to the inverse square of instantaneous distance from the compact object.   }  
  % results heading (mandatory)
  { The depth of the ionization zone depends on the illumination with photons above the ionization threshold, which is much higher for He than for H. Thus in the new model the emission distribution round the ring is expected to be more uniform for H$\alpha$ than for He I, under the supposition that these are recombination lines. The new model provides an excellent description of the observations, including H$\alpha$ intensity, and the variations of the H$\alpha$ and He I spectra with orbital phase are described quantitatively.}
 % conclusions heading  (optional), leave it empty if necessary 
{ While the description of the data provided by this ionization zone model is little different from that provided by an earlier model, the new model is more secure because it makes no appeal to \textit{ad hoc} decay parameters and locates the differences between H$\alpha$ and He I spectra in the different ionization potentials. If the circumbinary disk scenario is essentially correct, as now seems even more likely, the mass of the binary system must exceed 40 $M_\odot$ and the compact object must be a rather massive stellar black hole.}


   \maketitle
%
%________________________________________________________________

\section{Introduction}
\subsection{The microquasar SS 433}

  The Galactic microquasar SS 433 is unique in continually ejecting jets of relatively cool gas at approximately one quarter of the speed of light, particularly prominent in H$\alpha$ emission. The system is a binary with a 13 day orbital period and the photosphere of the accretion disk is very bright; the compact object is a super-Eddington accretor. The mass of the system exceeds 10$M_\odot$ and it is of considerable interest to establish the mass of the compact object. The optical spectra contain not only emission lines from precessing relativistic jets but also emission lines from gas motion within the system. The dominant contributions to these stationary emission lines come from a fast wind emitted preferentially perpendicular to the accretion disk and an additional component, displaying a persistent two horn structure of just the kind expected for emission from an orbiting ring, or a disk, seen more or less edge on. The horn separation corresponds to a rotation speed in excess of 200 km s$^{-1}$ and attributed to material orbiting the centre of mass of the binary implies a system mass in excess of 40 $M_\odot$. 

The H$\alpha$ spectra were originally discussed in  Blundell, Bowler \& Schmidtobreick (2008); departures from the pattern expected for a uniformly radiating ring are present and are more pronounced in He I emission lines.  The H$\alpha$ and He I spectra here addressed are displayed in Schmidtobreick \& Blundell (2006a), Fig. 2. It takes little more than inspection of that figure to extract certain salient features. First, the H$\alpha$ line has two narrow components with a separation that scarcely varies from JD 2453000 + 245.5 to + 274.5. In particular, there is no trace of motion of the kind expected if the ring orbited an object in turn orbiting the centre of mass of the binary. This is even clearer in Fig.1 of Blundell, Bowler \& Schmidtobreick (2008). Secondly, the red and blue horns of H$\alpha$ fluctuate in intensity in antiphase, although smooth periodicity is disturbed by occasional red outbursts. The epochs of equal intensity occur very close to orbital phases 0.5 and 1 (a phase of zero corresponding to mid primary eclipse) and the He I lines also pass through this symmetric configuration at the same time. In Fig.2 of Schmidtobreick \& Blundell (2006a) this is particularly clear for days +248.5 (orbital phase 0.51) and +274.5 (orbital phase 0.5). The He I emission lines fluctuate in intensity from red to blue with much greater amplitude than H$\alpha$ and the red horn is most dominant at orbital phase 0.75 (when the compact object is receding), the blue at phase 0.25 when it is approaching. 

   The obvious interpretation of these features is that both H$\alpha$ and He I emission are produced in a circumbinary ring (or inner rim of a disk) and are stimulated by radiation from the compact object. A particular problem with this picture is that the He I lines are emitted from a much more local patch of the ring than H$\alpha$. Before the present paper,the only explanation proffered was that any patch of material stimulated by radiation impact takes some time for the stimulated emission to die away - and that He I emission decays quicker than H$\alpha$. In the most recent version of such a model (Bowler 2011a) emission was stimulated proportional to the intensity of the stimulating radiation and stimulated material cooled with a time scale of less than 1 day in the case of He I, but several days for H$\alpha$ emitting regions.
   
   As a purely phenomenological treatment this model is extremely successful. The requirement of cooling times of the order of days is however a worrying feature, as is the difference of cooling times between He I emitting regions and H$\alpha$ regions in the same disk rim. The second point is obvious; the former less so. The problem is that both H$\alpha$ and He I emission lines most likely result from recombination in a highly ionized environment. The recombination timescales are only of the order of a day or greater for electron densities of $\sim$ 10$^{8}$ cm$^{-3}$ or less - several orders of magnitude below the densities likely to be found in the environs of SS 433. Electron densities $\sim$ 10$^{11.5}$ cm$^{-3}$ in the circumbinary disk have been inferred from Balmer decrements (Perez \& Blundell 2010).
   If it is assumed that H$\alpha$ and He I are recombination lines and emitted in a region where the electron density is maybe 10$^{11}$ cm$^{-3}$ or greater, then recombination times are negligible, reckoned in minutes rather than days. In the model of Bowler (2011a) the He I data are already close to this, for a decay time of 1 day is small on the scale of the orbital period. It is however evident that the emission of H$\alpha$ must be much more uniform round the disk rim than for He I and if recombination times are short the effective stimulation intensity for H$\alpha$ must be much more uniform than for He I. An explanation for this is needed and it has to be natural in terms of the physical differences between hydrogen and helium atoms. 
   
   \subsection{The ionization zone model}
   
   The most obvious such difference is in the ionization potentials, 13.6 eV for hydrogen and 24.6 eV for helium. Stimulation of ring material by impact of stellar wind or outflowing material is not likely to be affected by this difference, but hydrogen can be ionized by wavelengths $<912 $ \AA\ and helium $<504 $  \AA; most quasi-thermal sources of ultra violet and soft X-rays will emit a much lower flux of photons capable of ionizing helium than of ionizing hydrogen. In nebulae stimulated by O or B type stars, the radius of the He I Stromgren sphere is much less than that of the H II region for just this reason. I suggest that the difference between the H$\alpha$ and He I emission spectra in the stationary lines of SS 433 is due to an analogous effect. The idea is very simple. If there are not enough ionizing photons to maintain almost complete ionization in the disk, then the volume of ionized helium will expand and contract radially as the source in the compact object approaches and recedes. If however there are more than enough photons to ionize hydrogen at the distance of closest approach of the compact object, then the volume of ionized hydrogen will stay constant until the flux has fallen to maybe one half of the peak value, or even all the way round the ring. This would give much more uniform emission of H$\alpha$ round the rim of the disk and potentially explains the difference between H$\alpha$ and He I in terms of the simple physical differences between the parent atoms.
   
    There is another effect related to the different ionization potentials that is probably relevant. The companion is less bright than the environs of the compact object and presumably cooler: it might be too cool to ionize helium but hot enough to ionize a significant amount of antipodean hydrogen.

   In this note I implement this ionization zone model and show that it is even better at providing a description of the ring emission data than preceding models appealing to hot spots with different decay times. A major objection to models for ring emission (see the account in Bowler (2011a)) has been removed.

    \section{Implementing the ionization zone model}

    \begin{figure}[htbp]
\begin{center}
{ \includegraphics[width=9cm,trim=0 0 0 2]{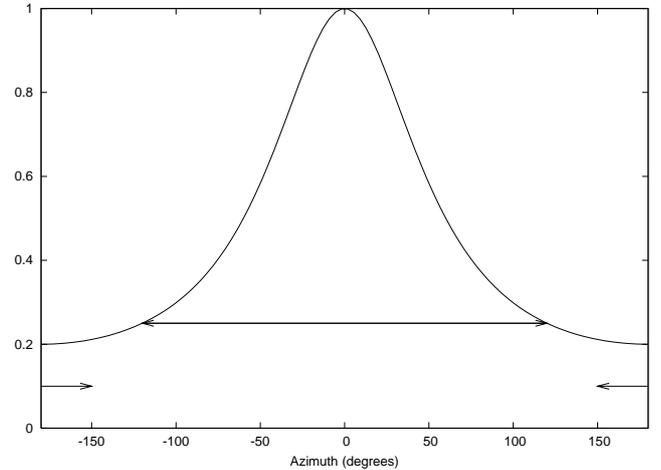} }        
   \caption{The curve shows the illumination of the circumbinary disk by an effectively point source located at the compact object. The azimuthal angle is measured from the line connecting the companion with the compact object; thus intensity 1 at azimuthal angle 0$^{\circ}$ corresponds to illumination of that patch of the circumbinary ring closest to the compact object. The illumination pattern rotates round the ring with a period of 13.08 days. The curve corresponds to a ring orbit radius of $1.6A$ and a mass for the compact object a little less than that of the companion.For He I the emission function was taken to be that of the illumination, apart from the eclipse range of 30$^{\circ}$ on each side of 180$^{\circ}$, where it was set equal to zero. For H$\alpha$ the illumination function was truncated initially above 0.25 (double ended arrow)to yield the emission, with a value of 0.1 in the eclipse zone (inward arrows; see text). 
   }
\label{fig:ideagram}
\end{center}
\end{figure}

\subsection{Structure of emission functions}

     I have assumed that at any moment the circumbinary ring is illuminated  proportional to the inverse square of the distance between the circumbinary ring material and the compact object. The model illumination as a function of  azimuthal angle relative to the compact object is displayed in Fig.1. The assumed geometry is for a circumbinary ring radius of $1.6A$ and a compact object orbit of radius $0.6A$, where $A$ is the semi-major axis of the binary. Because primary eclipses last approximately two days, illumination by the compact object is eclipsed out over a range of 60$^{\circ}$. 
     
     The emission function has been taken as identical in shape to illumination by the compact object for the case of He I lines, but because of the assumption that there is a superfluity of photons capable of ionizing neutral hydrogen, for H$\alpha$ the illumination function was truncated; one quarter of the maximum flux deemed sufficient to ionize all the hydrogen. In addition (suggested by the data) a level of 0.1 was allowed during eclipse, for the companion is less luminous than the photosphere of the compact object, yet may ionize hydrogen but not helium. In Fig.1 the truncation is indicated by the horizontal double headed arrow and the eclipse zone by inward pointing arrows. The two levels set were not fine tuned but adjusted until recognisable  shapes were obtained for the spectra. It is of some interest that the He I spectral shapes require the eclipse cut, or something close to it. Similarly the H$\alpha$ spectral shapes required some compensation for eclipse in addition to the rather flat emission function over the range of approximately -100$^{\circ}$ to +100$^{\circ}$ .The cuts for H$\alpha$ emission shown in Fig.1 apply to the model calculations illustrated in Figs. 6 \&7. They were arrived at by playing around and are probably not the most suitable, as discussed in section 3. 2. The sharp edges in the assumed distribution functions are not expected to occur in nature and are responsible for the fine structure to be perceived in the rotational and systemic velocities calculated from the model (Figs. 4 \& 5).

    \begin{figure}[htbp]
\begin{center}
  \includegraphics[width=6.38cm]{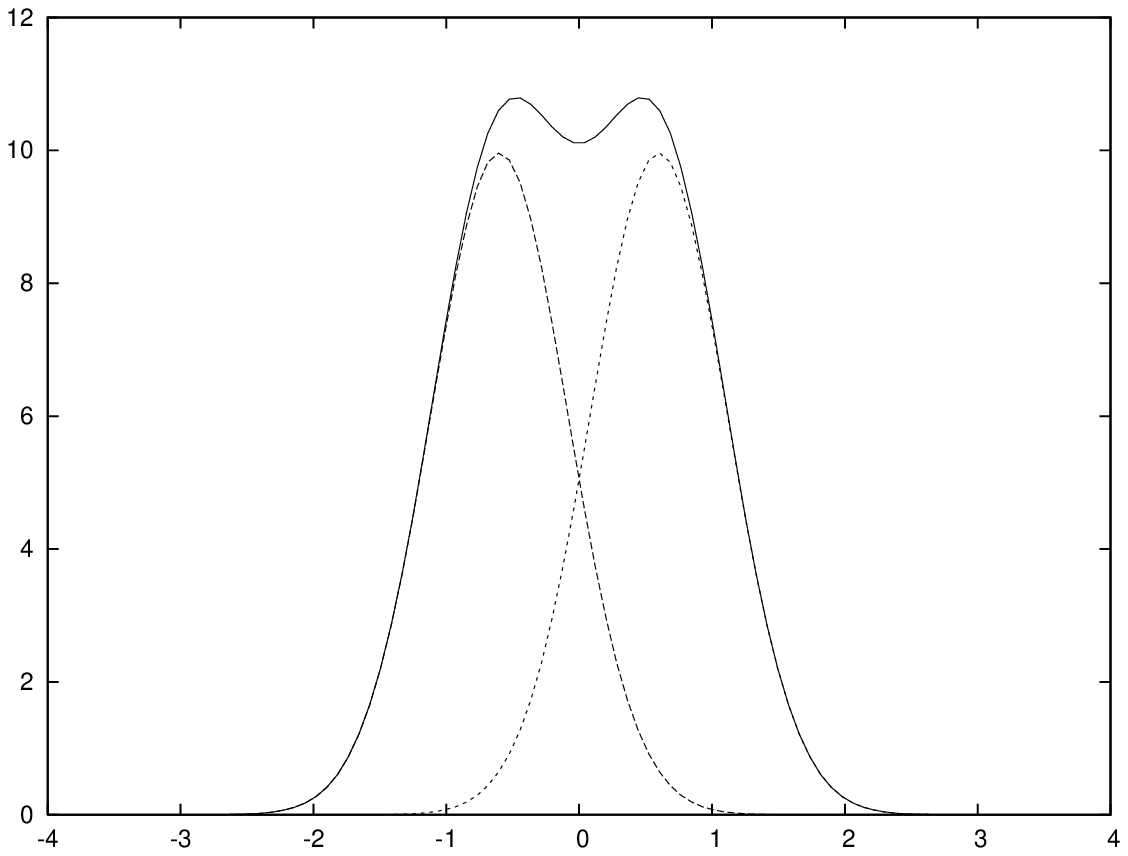}
  \includegraphics[width=6cm,trim=10 0 0 0]{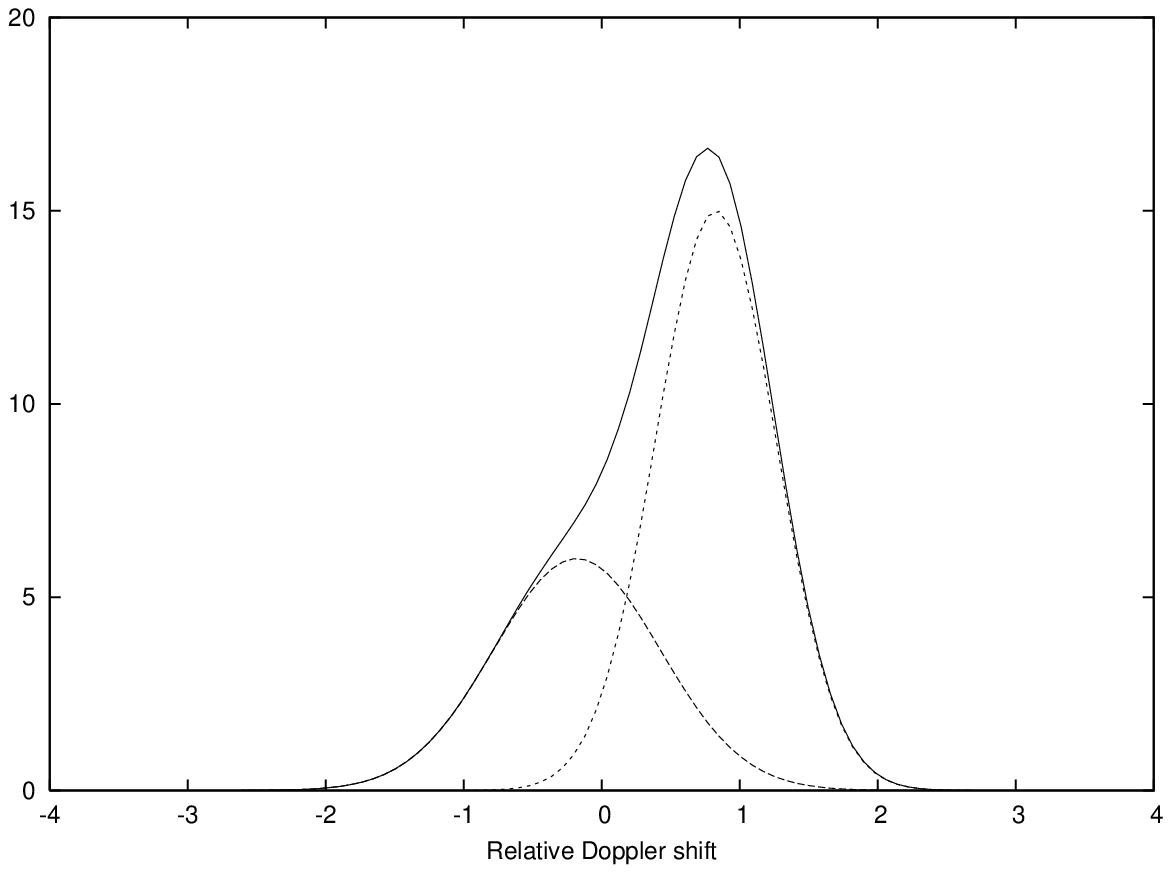}
\caption{ This is for the model He spectra. The \textit{upper panel} is for orbital phase 0.5 and the \textit{lower} for just after phase 0.75. As in the He data, the fitted components are further apart in the symmetric configuration. }
\label{fig:248}
\end{center}
\end{figure}

    \begin{figure}[htbp]
\begin{center}
   \includegraphics[width=6cm,trim=6 0 0 0]{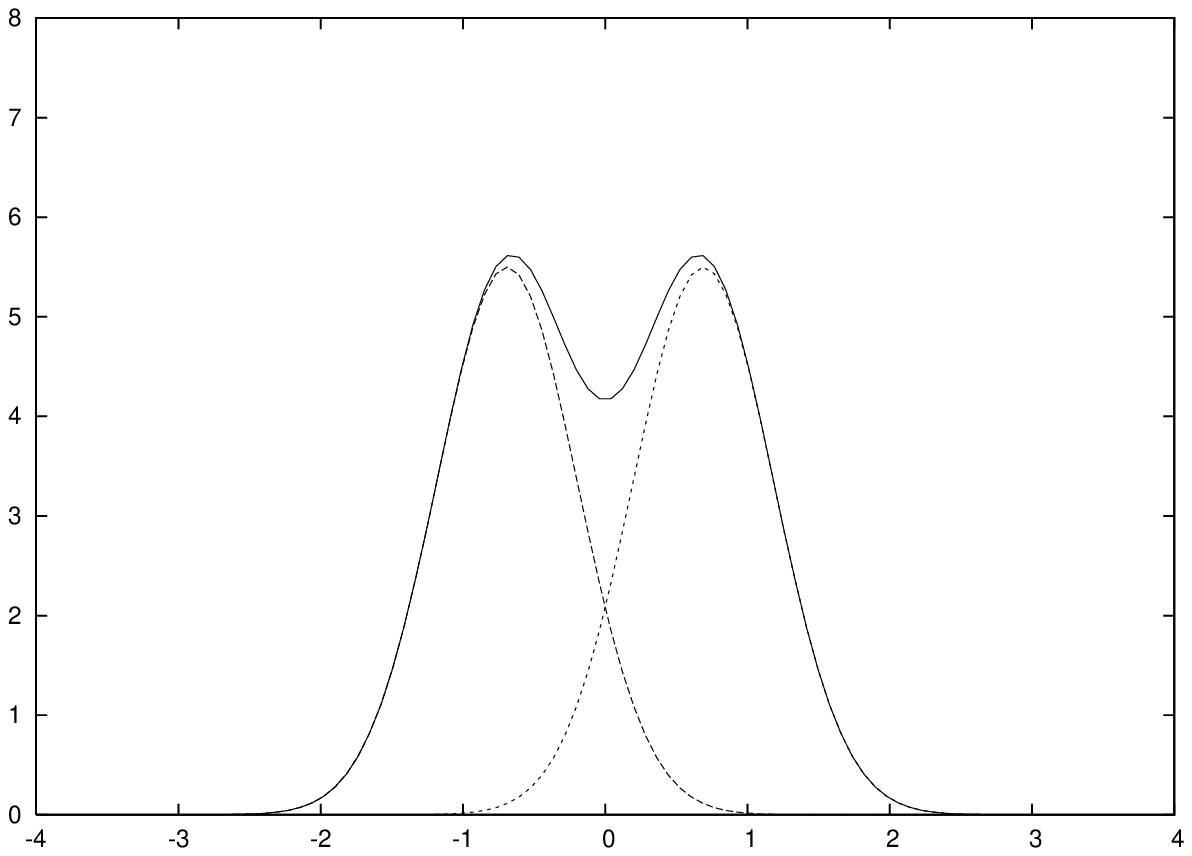}
   \includegraphics[width=6.3cm]{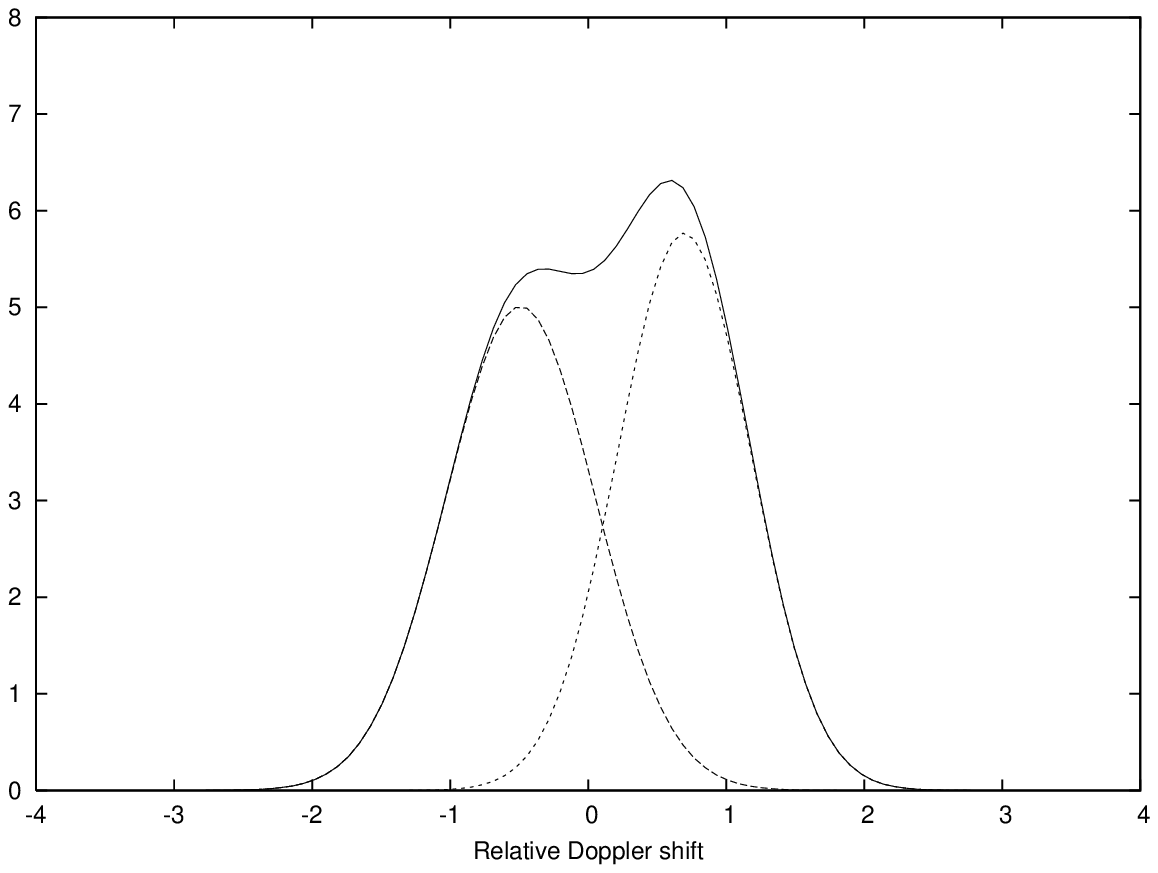}
\caption{ This figure illustrates spectra for the H$\alpha$ model and is to be compared with the model for He in Fig.2 The components are further apart for phase 0.5 (\textit{upper panel}), where the red and blue horns are of equal intensity than for phase 0.75 (\textit{lower panel}). This is in accord with Fig.2, but in the H$\alpha$ data it is the other way round. }
\label{fig:disc}
\end{center}
\end{figure}

    \subsection{Spectral shapes}
    
      As in Bowler (2010b, 2011a), I generated spectral shapes for orientations relative to the line of sight every $20^{\circ}$, as a function of the variable $x$, the red shift as a fraction of the maximum possible. Thus a value of $x=-1$ corresponds to radiation from material in an orbiting ring approaching tangential to the line of sight. The generated spectra were then decomposed into a superposition of two Gaussian functions; examples are shown in Fig. 2 (for He I) and Fig. 3 (for H$\alpha$). These model spectra may be compared with the sequences of spectra in Fig. 2 of Schmidtobreick \& Blundell (2006a). In all illustrations of the results of implementing the ionization zone model I have assumed a rotation speed of 250 km s$^{-1}$, corresponding to $x=1$ and systemic recession of 70 km s$^{-1}$.
      In Fig. 4 of Bowler (2011a) I showed the sequence of model He I spectra over a full orbit, in the same format as the compilations of data shown in Fig. 2 of Schmidtobreick \& Blundell (2006a). The revised model presented in this paper generates almost identical shapes, because the supposed cooling or decay time for He I in Bowler (2011a) was only about a day and hence had little effect other than to introduce a slight delay of that order; compare Fig. 2 with Figs. 2 \& 3 of Bowler (2011a).

 \section{Results of matching the new model to the data}
 
 \subsection{He I spectra}
    The Doppler speeds of the narrow Gaussian components fitted to the spectra in the data and to the model spectra may be compared in various ways. Here, I carry out the comparison in terms of the mean of the Doppler shifts and, separately, half the difference. For a uniformly glowing ring the former would be the systemic velocity and the latter the rotation speed of the orbiting ring. Fig. 4 shows the variation of the apparent rotational speed with time and Fig. 5 the apparent systemic recession for the He I 6678 \AA\ data, day by day from JD +245 to +274.  It is important to note that the phase of the model oscillations is not an adjustable parameter and the agreement with the data is evidence that the hottest spot on the circumbinary ring rotates with the phase of the compact object. The hottest spot is given by the line between the companion and the compact object, extrapolated to the ring, and lies on the line of sight when the compact object and disk eclipse the companion (orbital phase 0.5); JD + 248.36  according to the Goranskii ephemeris (Goranskii et al 1998). The data are the same as in Bowler (2011a) and the model results very similar, because of the unimportant role of a cooling time of about one day there assumed. The absence of this smearing factor is responsible for the delicately sculpted fine structure in Figs. 4 \& 5 and has also resulted in a small shift of the coarse structure to earlier times. The structures in the data lag the model by possibly as much as a day; otherwise the agreement is as good as could be expected.

\begin{figure}[htbp]
\begin{center}
  \includegraphics[width=9cm,trim=0 0 0 140]{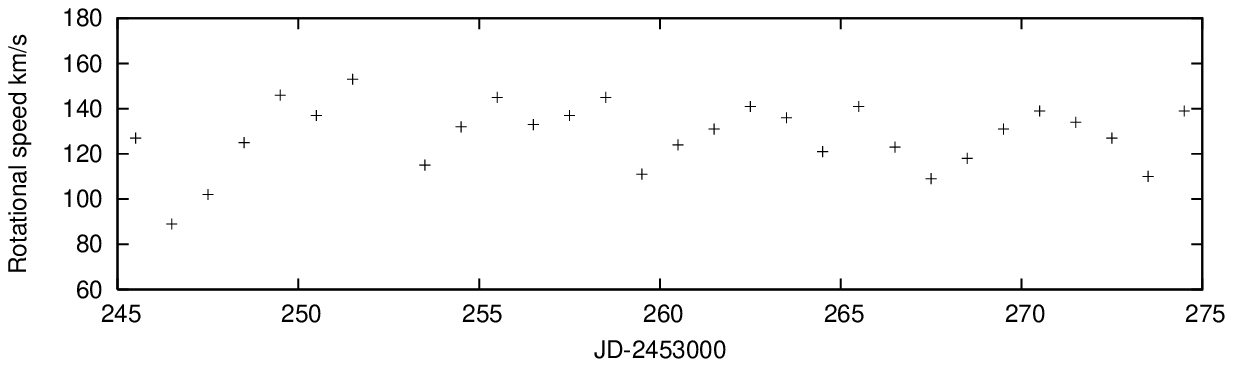}
  \includegraphics[width=9cm,trim=0 0 0 140]{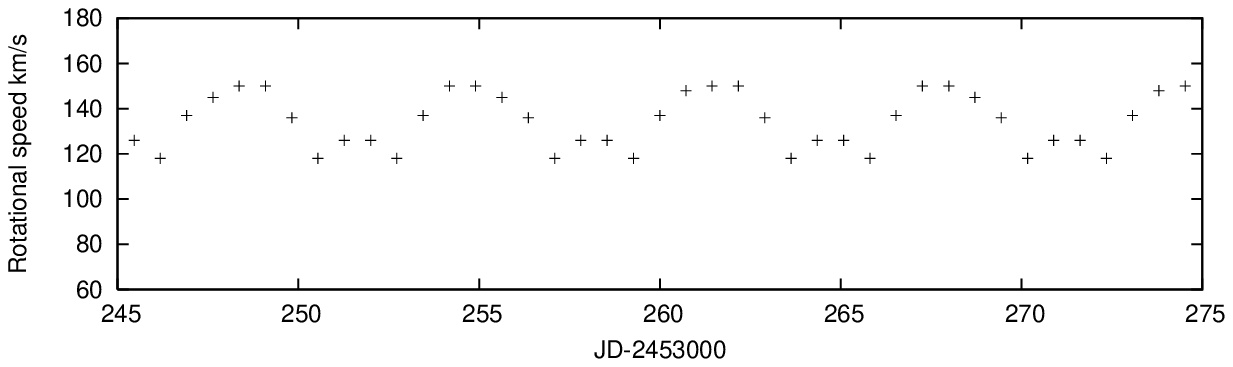}
   \caption{ The nominal rotational velocity of the circumbinary disk, as obtained from the differences between the red and blue components. The \textit{upper panel} is for He I 6678 \AA\ and the \textit{lower panel} is the model calculation. The data could lag the model by as much as 1 day.}
\label{fig:timesequence}
\end{center}
\end{figure}

\begin{figure}[htbp]
\begin{center}
  \includegraphics[width=9cm,trim=0 0 0 140]{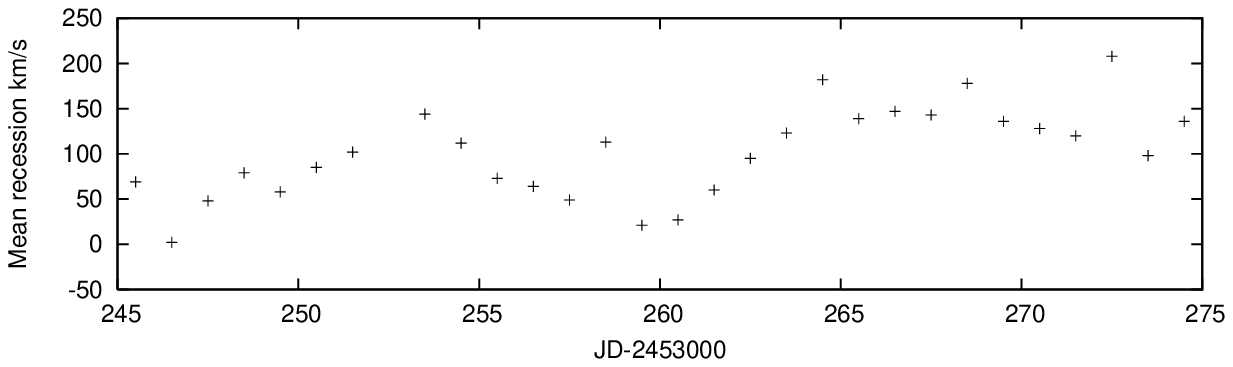}
   \includegraphics[width=9cm,trim=0 0 0 140]{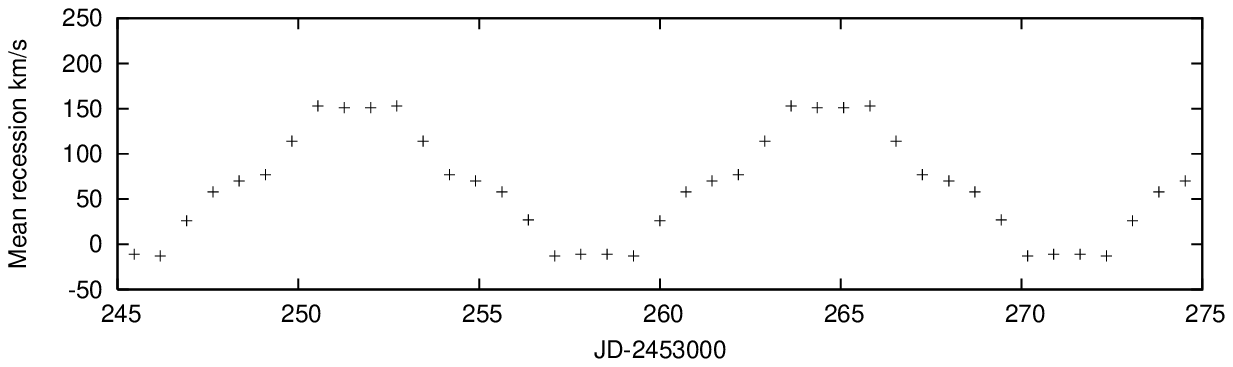}
   \caption{ The mean recessional velocity of the red and blue circumbinary disk components as a function of time. The \textit{upper panel }is for He I 6678 \AA\ and the \textit{lower panel }is the model calculation. The data could lag the model by perhaps $1/2$ a day.}
\label{fig:timesequence}
\end{center}
\end{figure}

    \subsection{H$\alpha$ spectra}
The structure of the model for H$\alpha$ emission is significantly different from any of the previous models. In the absence of any cooling time effects the emission function is symmetric about the line joining the compact object and its companion. However, with a fairly flat emission function as implemented, the component lines again run railroad straight; the data (from Blundell, Bowler \& Schmidtobreick 2008) are compared with the ionization zone model in Figs. 6 \& 7.
The comparison between the rotational speed of the ring extracted from the H$\alpha$ data and the model is of peculiar interest. This is displayed in Fig. 6. The topmost panel is the data and although there appear to be periodic minima the statistical significance is perhaps marginal. The middle panel displays the results from the model as described in section 2. The oscillations are of comparable size, with amplitude of $\sim$ 15 km s$^{-1}$ but they are not in phase with those of the data. This is presented vividly in the bottom panel of Fig. 6, where the two samples are braided. The curves they describe osculate at days +248, 255, 262 and come close at 268 and even 274. These dates are separated by half an orbital period and +248 has orbital phase 0.5 (compact object closest to us) and +255 orbital phase 0 (or 1), primary eclipse. In both the data and the model these dates correspond to symmetric configurations, where the red and blue horns are of equal magnitude, for both the He I and H$\alpha$ spectra. The two curves defined are in almost perfect anti-phase. Thus the gross features of the spectra are the same in both data and model but there is fine structure not matched. In the data the separation between the red and blue horns for the symmetrical configuration is a little smaller than the separation when the horns are skewed. In the model as illustrated in Fig.1, the separation is greater for the symmetric configuration than for asymmetry. This discrepancy requires a hotter spot in the antipodes and is removed by a different choice of parameters in the model of Fig.1. In that model the emission function from the compact object was truncated at 0.25 of the maximum value and the eclipse zone filled in to the level of 0.1 by radiation from the companion. If the truncation is performed at a level of 0.5 and the eclipse zone filled in to the level of 0.4, then the pattern for the model shown in Fig. 6 b, c is shifted by just over 3 days, or half a period of the oscillations in rotational velocity, and hence in phase with the data. The amplitude is a little smaller but in no way inconsistent with the data. A slightly more elaborate model involves attributing to the companion an illumination function like that of the compact object, but with a maximum intensity (for that part of the ring closest to the companion) 0.4 of the maximum compact object luminosity. If the emission caused by the compact object is truncated at 0.8 of the level corresponding to total absorption of H ionizing photons, similar results are obtained.

  All of these versions of the underlying model generate spectra qualitatively similar to each other and to the data and the rotation velocity minima seen in the H$\alpha$ data are matched without spoiling the qualitative agreement with the shapes of individual spectra if the eclipse zone is sufficiently filled in by the companion. While a fast wind from the companion impacting on the circumbinary ring might achieve this, it is not plausible that impacts at speeds $\sim$ 100 km s$^{-1}$ and greater would stimulate H$\alpha$ and not He I. Radiation from the companion could do this, because of the higher ionization potential for helium. Thus the best description of the H$\alpha$ spectra is provided by a combination of the two effects of the difference in ionization potentials discussed in section 1.2. For H$\alpha$, where intensities vary comparatively little with phase, it is detail in the emission function that determines whether the symmetric configuration corresponds to a maximum or minimum in the nominal orbital velocity.  There seems little point in further crafting an emission function for H$\alpha$ that inverts the pattern deriving from Fig.1 - it would be arbitrary and motivated only by a desire to achieve a spurious perfection. 
 
    Fig. 6, particularly the lowest panel, illustrates vividly that the symmetric configurations correspond in both cases to extremes of the nominal rotational velocity from H$\alpha$ and that the two curves are essentially in perfect anti-phase; they are $\pi$ out of phase and not ${\pi}/2$. The H$\alpha$ and He I emission lines have their origin in orbiting material and not in a ring expanding at 250 km s$^{-1}$.
     Fig. 7 is included largely for completeness. It shows the mean recessional velocity of the two horns for both H$\alpha$ and the original model of Fig.1; the predictions of the other versions are not significantly different here. The excursions are much smaller than for He I (Fig. 5), where the dominant emission swings much more violently from red to blue. The H$\alpha$ lines are railroad straight in comparison.

  \begin{figure}[htbp]
\begin{center}
  \includegraphics[width=9cm,trim=0 0 0 140]{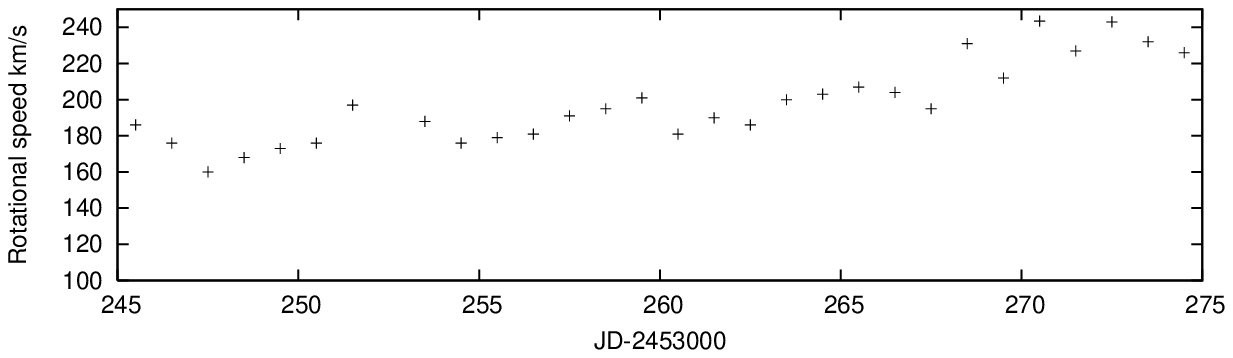}
  \includegraphics[width=9cm,trim=0 0 0 140]{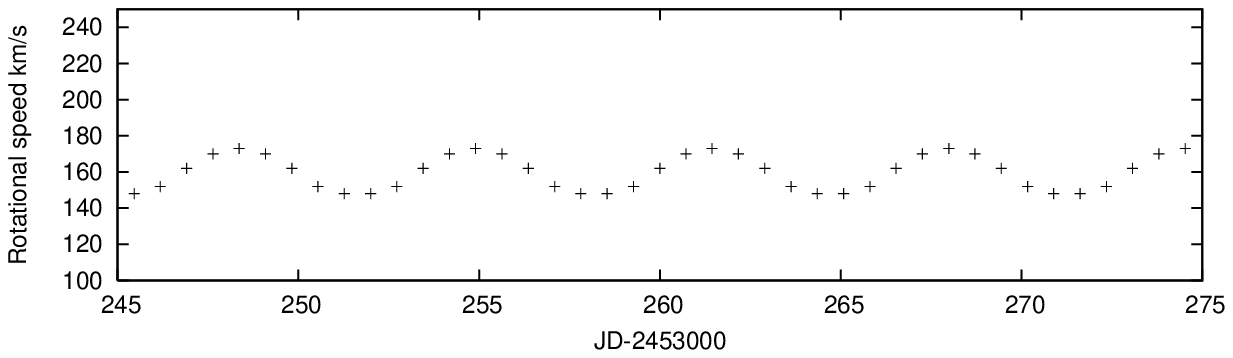}
  \includegraphics[width=9cm,trim=0 0 0 140]{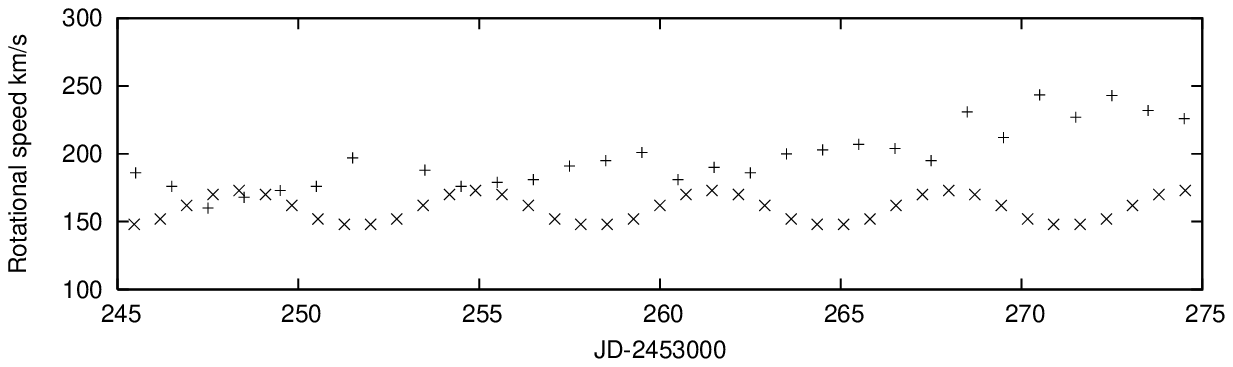}
   \caption{ The nominal rotational velocity of the circumbinary disk, as obtained from the differences between the red and blue components. The \textit{upper panel }is for H$\alpha$ and the \textit{middle panel} is the model calculation. The anti-phase is noticeable, emphasised by the simultaneous plotting of the two in the \textit{bottom panel}; see text. }
\label{fig:timesequence}
\end{center}
\end{figure}

\begin{figure}[htbp]
\begin{center}
  \includegraphics[width=9cm,trim=0 0 0 140]{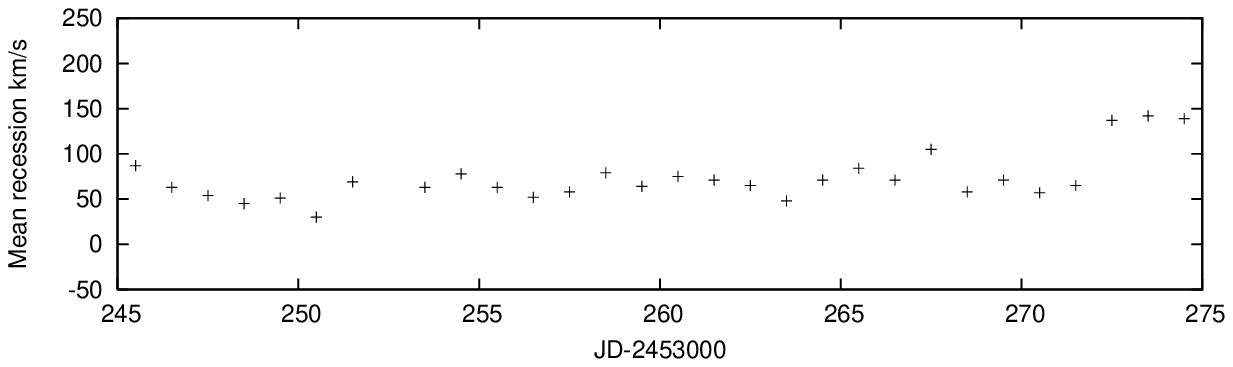}
   \includegraphics[width=9cm,trim=0 0 0 140]{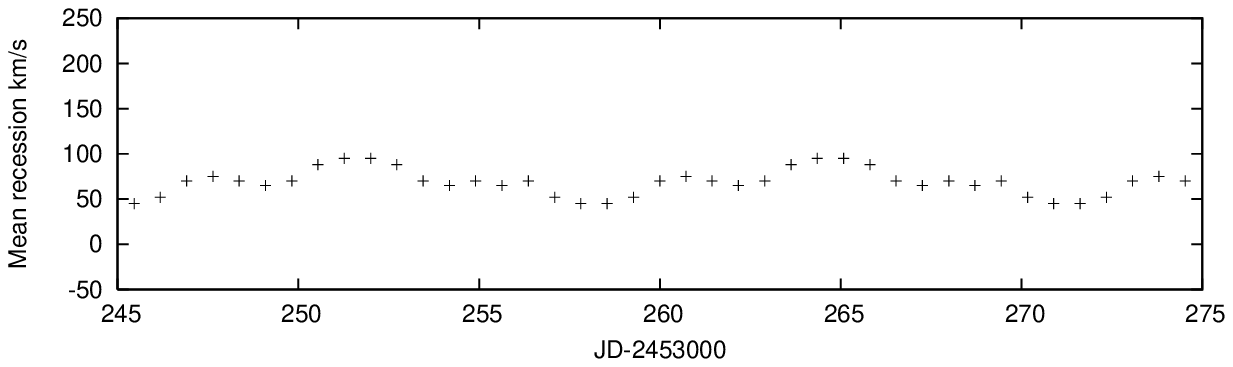}
   \caption{ The mean recessional velocity of the red and blue circumbinary disk components as a function of time. The \textit{top panel} is for H$\alpha$ and the \textit{bottom panel} is the model calculation. The wiggle is much smaller than for He I. }
\label{fig:timesequence}
\end{center}
\end{figure}

 \section{Matters arising} 
 
 \subsection{Physical dimensions of the ring}
 There is a very important point that has not been dealt with so far. For the ionization zone model to be acceptable it is necessary for the thickness of the circumbinary ring to be realistically plausible. The depth ${d}$ of the fully ionized zone for hydrogen can be simply estimated by equating the incoming flux of photons above 13.6 eV with the number of decays per second in the volume $d$. The flux can be obtained from the quantity known in the trade as $Q(H^0)$ which is the effective number of hydrogen ionizing photons per second emitted by the corona of the accretion disk. This is estimated as $\sim$ 10$^{48}$ s$^{-1}$ for an ultraviolet spectrum similar to that of a BO or O9 star, spectra sufficiently soft that the depth of a zone of fully ionized helium is substantially less than that of fully ionized hydrogen (Osterbrock \& Ferland 2006; see also Fabrika 1993. Fabrika 1993 considered the possibility of H$\alpha$ emission from a disk external to the binary system through recombination in an essential fully ionized environment. He, however, seems to have had in mind excitation of a radially expanding spiral - out to very large distances - rather than an orbiting circumbinary disk. ). For electron density $\sim$ 10$^{11}$ cm$^{-3}$ the depth of a fully ionized hydrogen zone starting at a distance of $\sim$ 10$^{13}$ cm from the compact object is $d \sim$ 10$^{12}$ cm. The radius of the circumbinary ring is approximately twice the semi-major axis of the binary $A$ (which is $A \sim$ 5 10$^{12}$ cm for the system mass inferred from the circumbinary ring) and the distance from the compact object to the inner edge varies from approximately $2.5A$ to $1.5A$. These numbers all seem perfectly reasonable and the width of the ring, through which hydrogen ionizing photons punch over much of the azimuthal range, is about one tenth of the radius. If it is assumed that the ring is approximately toroidal in form, the mass of gas in such a ring is only $\sim$ 10$^{-8}$ $M_\odot$ (far too small to account for an orbital precession taking 100 years, as conjectured in Bowler (2010a)) and we can estimate the intensity of H$\alpha$ from the circumbinary ring.
 
 \subsection{Intensity radiated in H$\alpha$}

   The model is sufficiently circumscribed that a realistic estimate of the intensity of H$\alpha$ radiation can be made. Since such a ring intercepts a substantial fraction of H ionizing radiation produced by the environs of the compact object, which has some of the characteristics of an O type star, it is not surprising that there is good agreement. The ring intercepts approximately 10$^{47}$ such photons each second and this is the (approximate) number of recombinations. The total H$\alpha$ luminosity from the ring is then $\sim$ 10$^{35}$ erg s$^{-1}$. Allowing for the distance to SS 433 and interstellar extinction with $A_V$ $\sim$ 8 the H$\alpha$ flux entering terrestrial telescopes is $\sim$2.5  10$^{-13}$ erg cm$^{-2}$ s$^{-1}$,  or 2.5 10$^{-16}$ W m$^{-2}$. This is spread over a range of approximately 10 \AA\ , leading to a predicted peak intensity of 2.5 10$^{-14}$ erg cm$^{-2}$ s$^{-1}$ \AA\ $^{-1}$,  or 2.5  10$^{-13}$ W m$^{-2}$ $\mu$m$^{-1}$. There is a direct measurement to be found in Fig.2 of Perez \& Blundell (2010). The contribution of the circumbinary ring was there fitted by a single gaussian and the peak height found to be 10$^{-13}$ W m$^{-2}$ $\mu$m$^{-1}$, approximately 5 times the continuum background.  The data of Schmidtobreick \& Blundell (2006 a, b) are not calibrated for intensity but the two horns interpreted here as coming from the circumbinary disk have standard deviations of a few \AA\ and peak heights about 5 times the nearby continuum. For the same continuum intensity recorded in the spectrum of Perez \& Blundell (2010) the inferred intensity of radiation from the circumbinary disk is thus about the same and both somewhat less than my estimate above. If the continuum background in the 2006 data is more typically $\sim$10$^{-14}$ erg cm$^{-2}$ s$^{-1}$ \AA\ $^{-1}$ then those data imply an intensity of H$\alpha$ from the ring a few times larger than my estimate. Thus the ionization model presented has no trouble in accounting for such measured intensities as exist. The precise number depends on uncertain details; for example, the predicted intensity would be increased by a factor of 4 if the flux of hydrogen ionizing photons were increased by a factor of two and the transverse dimensions of the radiating torus by the same factor.
  
  \subsection{Radiation from the wind and the jets}
 
 The next matter arising is the possible relation between the mechanisms underlying decay of radiation from packets of gas in the circumbinary ring on the one hand and in the wind above the accretion disk on the other. In this discussion of the ring, any given packet of gas fades in H$\alpha$ or He I emission as a result of a reduction of ionizing radiation in a region of constant density. The H$\alpha$ radiation from any packet of gas in the wind seems to fall away with a characteristic time of a few days after escape (Bowler 2011b). The density falls as the material blows outwards, but its ionization is expected to be independent of distance from the compact object, once both the gas density in the wind and the flux of ionizing photons follow inverse square laws (Fabrika 1993; see also Osterbrock \& Ferland 2006.) The recombination rate in such a packet of wind thus falls inversely with the square of distance from the ionizing source. If escape is achieved at a distance of a few photospheric radii and line emission there commences, emission has fallen by a factor of $\sim$ 10 after about 10$^{13}$ cm. At a speed of 1000 km s$^{-1}$ this corresponds to just over a day; two days if a more appropriate speed is 500 km s$^{-1}$. This crude estimate is a bit short, but the decay time of wind packets may also fit into the ionization zone framework - it  is not obvious what other kind of mechanism could be responsible. The most famous source of fading H$\alpha$ emission in SS 433 is of course found in the bolides ejected at 0.26$c$ in the precessing jets. Ionizing radiation from the nozzle has been considered as the machinery for maintaining radiation, but there is also the possibility of interaction of the bolides with the wind from the disk; see Fabrika (2004) for a review.
 
 \subsection{Paschen lines from recombination}
 
 The final  matter arising is a potential embarrassment. The higher order Paschen lines in the hydrogen spectrum also exhibit the two horned structure characteristic of emission from ring seen more or less edge on and with a separation corresponding to rotation at about 200 km s$^{-1}$ (Filippenko et al 1988, Schmidtobreick \& Blundell 2006b). Schmidtobreick \& Blundell have not published their daily Paschen spectra (Pa-$\kappa$ to Pa-$\nu$), but it is remarked in Blundell, Bowler \& Schmidtobreick (2008) that the Paschen spectral shapes behave more like He I than H$\alpha$. If the Paschen emission lines also result from recombination in a zone of more or less complete ionization, the Paschen lines might be expected to be produced with the azimuthal uniformity of H$\alpha$, yet their stimulus appears not to punch through the ring even at maximum illumination. It might work like this. The photo-ionisation cross section falls very rapidly as the photon energy rises above threshold, so the relatively low energy photons penetrate less and the higher energy photons penetrate deeper into the ring material. The lower energy photons produce low energy electrons and high energy photons produce high energy electrons. High order Paschen lines are generated via recombination to H states of high principal quantum number $n$ and the capture cross section falls much more rapidly with electron energy for high $n$ than for the lower values that can feed H$\alpha$ (see for example Searle 1958). There is thus reason to suppose that the high Paschen lines are preferentially produced nearer the inner rim of the circumbinary disk than H$\alpha$, which can be generated readily by deeply penetrating high energy photons.

  \section{Discussion}
  
  The sequence of spectra published in Blundell \& Schmidtobreick (2006a) have, qualitatively, all the characteristics of emission of H$\alpha$ and He I from a circumbinary ring, stimulated by the intense radiation from the compact object. The material is orbiting the binary and the data are not consistent with radially expanding ring structures, although a small radial component cannot be ruled out. The ionization zone model discussed in this note lifts the description beyond the merely phenomenological, providing an explanation of the differences between the H$\alpha$ and He I spectra in terms of the difference in the ionization potentials. The difficulties with purely phenomenological models, listed in Bowler (2011a), have largely disappeared.
  
  A circumbinary ring orbiting at $\sim$ 250 km s$^{-1}$ implies a system mass of $\sim$ 40 $M_\odot$ and an orbital velocity for the companion in excess of 130 km s$^{-1}$ (Blundell, Bowler \& Schmidtobreick 2008). Two sequences of absorption spectra in the blue have suggested that the orbital speed of the companion is 58 km s$^{-1}$ (Hillwig \& Gies 2008, Kubota et al 2010). It would be agreeable if the data on the circumbinary disk could be reconciled with a companion orbital speed of almost 60 km s$^{-1}$ but I have been quite unable to devise a suitable scheme. The observations themselves are not inconsistent if the spectra attributed to absorption in the atmosphere of a mid A-type companion were in fact produced by absorption of light from the companion in the circumbinary disk itself (Bowler 2010b, c). It is worth listing here a number of grounds for being cautious about associating the absorption spectra with the atmosphere of the companion, despite the consistency of observation with expectation (Hillwig \& Gies 2008). First, observations of absorption spectra in the blue have been made at a variety of different precession phases (not including the magic phase corresponding to maximum opening angle of the accretion disk) by Barnes et al (2006). The spectra look clean and look rather like mid A-type stars but they are not phased as they would be if they came from the companion. They are probably formed in the wind from the disk (Barnes et al 2006; see also Bowler 2011b) and at any rate there is agreement that not all the absorption features in the blue are formed in the atmosphere of the companion (Barnes et al 2006, Hillwig \& Gies 2008). There is material loose in the system that has absorption spectra like those of A- type stars, but ain't.
  
  The absorption spectra phased with the companion that imply an orbital speed of $\sim$ 60 km s$^{-1}$ were only observed for precession phases where the accretion disk is most open and spanned roughly the orbital phases 0 to 0.3 - from just before primary eclipse until just over a quarter of a period later. This interval was chosen to make the companion most visible and this choice would also give the best opportunity for seeing light from the companion being absorbed in material other than its own atmosphere, such as the circumbinary disk.
  The spectra analysed by Cherepashchuk et al (2005) were also taken with the accretion disk wide open and the Doppler shifts they observed are also phased as for production in the atmosphere of the companion and match a sine curve with period 13 days. The spectral resolution is inferior to the later observations but if absorption was in the atmosphere of the companion an orbital speed of 132 km s$^{-1}$ is implied, with an error of 9 km s$^{-1}$. Yielding a mass for the compact object of 18 $M_\odot$, this would not conflict with inferences from the circumbinary disk. The environment in the outer reaches of the system is complicated and it might be that there are absorptive regions more closely associated with the companion, perhaps even within its atmosphere. Different absorptive regions evidently dominate at different times. It might be that a serendipitous sequence of spectra could be companion dominated and it might even be that the data of Cherepashchuk et al (2005) constitute such a sequence. I am not aware that those data or the inferences from them have been formally retracted, but they do seem to have been universally ignored.
  
\section{Conclusions}

The model presented in this note explains the difference between the H$\alpha$ and He I spectra in the stationary lines of SS 443 in terms of a physical mechanism of different ionization zones due to different ionization potentials. It accounts for the radiated H$\alpha$ intensity and has removed the $ad$ $hoc$ aspects of earlier phenomenological models;  strengthening even further the evidence that the two horned structures are indeed signatures of a circumbinary disk orbiting at approximately 250 km s$^{-1}$. I know of no observations indubitably at odds with a system mass of approximately 40 $M_\odot$, but companion orbital velocities in the region of 60 km s$^{-1}$ do not seem reconcilable with those aspects of stationary emission spectra interpreted here and elsewhere as arising from a circumbinary disk stimulated by intense radiation from the accretion disk. The system mass is in the region of 40 $M_\odot$ and the compact object is a rather massive stellar black hole, just as argued in Blundell, Bowler \& Schmidtobreick (2008).

\end{document}